\documentclass[aps,preprintnumbers,amsmath,amssymb,twocolumn]{revtex4}

\usepackage{graphicx}
\usepackage{dcolumn}
\usepackage{bm}
\usepackage{color}
\usepackage{amssymb}   

\begin{document}

\title{Flow induced crystallization of penetrable particles}

\author{Alberto Scacchi}
\author{Joseph M. Brader}

\affiliation{
Department of Physics, University of Fribourg, CH-1700 Fribourg, Switzerland\\
}


\begin{abstract}
For a system of Brownian particles interacting via a soft exponential potential we investigate 
the interaction between equilibrium crystallization and spatially varying shear flow.
For thermodynamic state points within the liquid part of the phase diagram, 
but close to the crystallization phase boundary, we observe that imposing a 
Poiseuille flow can induce nonequilibrium crystalline ordering in regions of low shear 
gradient.
The physical mechanism responsible for this phenomenon is shear--induced particle migration, 
which causes particles to drift 
preferentially towards the center of the flow channel, thus increasing the local density 
in the channel center. 
The method employed is classical dynamical density functional theory.  
\end{abstract}

\maketitle

\section{Introduction} 

The creation of a colloidal crystal generally proceeds via the slow process of nucleation and 
growth within a dense, crowded environment \cite{oxtoby}. 
Technological applications requiring periodic order on long length scales (e.g.~photonic crystals) 
are hindered by the defects and grain boundaries which inevitably develop during 
the growth process. 
It is thus desirable to identify ways by which external fields can be employed to control 
the nucleation dynamics and thus optimize the quality of the resulting crystal. 
One way to achieve this aim is to drive the system out of equilibrium using mechanical deformation 
of the sample.  

Since the pioneering light scattering investigations 
of Ackerson and Pusey~\cite{pusey} it has been known that the judicious application of an 
oscillatory shear strain can induce three-dimensional crystalline order in colloidal 
fluids. 
In a similar spirit, Besseling {\it et al.} have performed real-space confocal microscopy 
experiments on colloidal mixtures under oscillatory shear~\cite{besseling}. 
Crucially, in both cases the bulk density of the experimental sample lies below that of the equilibrium 
freezing transition, such that these shear--induced crystals represent true out-of-equilibrium 
states; the microstructure relaxes back to equilibrium following the cessation of the flow.  
Although Besseling {\it et al.}~\cite{besseling} supported their experimental findings with simulation data, 
there currently exists no first-principles theoretical approach to address this phenomenon. 
In a recent study, Peng {\it et al.} have used molecular dynamics to study the influence of 
steady simple shear on the crystal growth kinetics~\cite{voigtmann}. However, in contrast to 
Refs.~\cite{pusey,besseling}, the thermodynamic statepoints considered in Ref.~\cite{voigtmann} 
were all in the crystal region of the phase diagram.   

 The aforementioned studies of flow--induced crystallization~\cite{pusey,besseling} considered 
simple shear flow, where the (time-dependent) 
shear gradient is constant in space. However, there are many situations of interest for which 
the shear gradient is a function of position.
For example, in the commonly encountered case of Poiseuille flow along a channel the shear 
gradient is a linear function of the distance from the channel center. 
It is well--known that when the shear--rate varies significantly on the scale of a particle 
diameter, then the particles will begin to exhibit a biased diffusion towards regions 
of low gradient: shear--induced migration \cite{LA}. 
The physical origin of this effect is that the collision frequency of a given particle with it's 
neighbours is not isotropically distributed over the surface of the particle; surface regions 
subject to a higher shear--rate will experience, on average, a greater number of collisions than 
those regions subject to a lower shear-rate \cite{MSB}. 
In the case of Poiseuille flow this mechanism leads to an increase in density 
at the centre of the channel. 

In this paper we will explore how the particle migration induced by Pouseuille 
flow can interact with the underlying equilibrium free energy to generate 
nonequilibrium ordered states in which a crystalline region appears at the center of the channel. 
In addition to presenting a novel kind of nonequilibrium state arising from the coupling 
of shear--rate gradients to free energy minima, we can imagine that these nonequilibrium 
steady states may be relevant for particle transport in certain microfluidic devices. 
The method we employ is dynamical density functional theory, modified along the lines 
of Refs.~\cite{KB1,KB2,SKB} to incorporate the non-affine particle motion necessary 
to capture particle migration effects.

The paper will be organized as follows: in section II we explain the model, in section III we give an overview of the employed  theory, in section IV we report our results and finally, in section V, we provide comments and an outlook for future work.

\section{Model}\label{pen}
In order to investigate the process of shear--induced crystallization we choose to employ a simple, 
minimal model of penetrable particles, namely the generalized exponential model (GEM). Within 
this model the purely repulsive pair interaction potential has an exponential form, with an 
exponent which can be adjusted to tune the strength of the repulsive force between particles. 
The continuous nature of the potential makes it particularly well suited for numerical studies.  
We consider the GEM model with an exponent of $8$, described by the pair-potential
\begin{equation}
\phi(r)=\epsilon \exp\left(-\left(\frac{r}{R}\right)^8\right),
\end{equation}
where $R$ and $\epsilon$ set the length and energy scales, respectively. 

Ordering phenomena in GEM models, especially concerning quasi-crystalline order \cite{andy_quasi_crys_1,andy_quasi_crys_2}, solidification \cite{andy_solidification}, generation of defects and disorder from quenching \cite{andy_quenching}, crystallisation under confinement at interfaces and in wedges \cite{Archer_GEM} have been recently addressed. 
An important feature of penetrable particle models, such as the GEM-8, is that at high densities 
they exhibit a so--called `cluster crystal' phase. Cluster crystals differ from more familiar 
hard--sphere type crystals in that each density peak can contain, on average, more than one 
particle. While the multiple occupancy of a given lattice site costs a potential energy of 
order $\varepsilon$ per particle pair, the resulting gain in entropy dominates to minimize the 
grand potential. 
This mechanism makes the study of crystallization in penetrable particle models more subtle 
than the case of simple hard spheres, for which only entropy plays a role. However, 
penetrabel particle systems have the significant benefit that simple mean field approaches 
can yield accurate results.


We consider a two-dimensional 
GEM-8 fluid system confined between a pair of parallel repulsive walls. 
We choose to work in two dimensions as this is the minimal situation in which 
crystallization can be studied and reduces the numerical demands of integrating 
the DDFT equations - three dimensional 
calculations would be very expensive. 
Moreover, the majority of the literature concerning the equilibrium GEM-8 model 
addresses the two-dimensional system.
\begin{figure}[t!]
\includegraphics[scale=0.45]{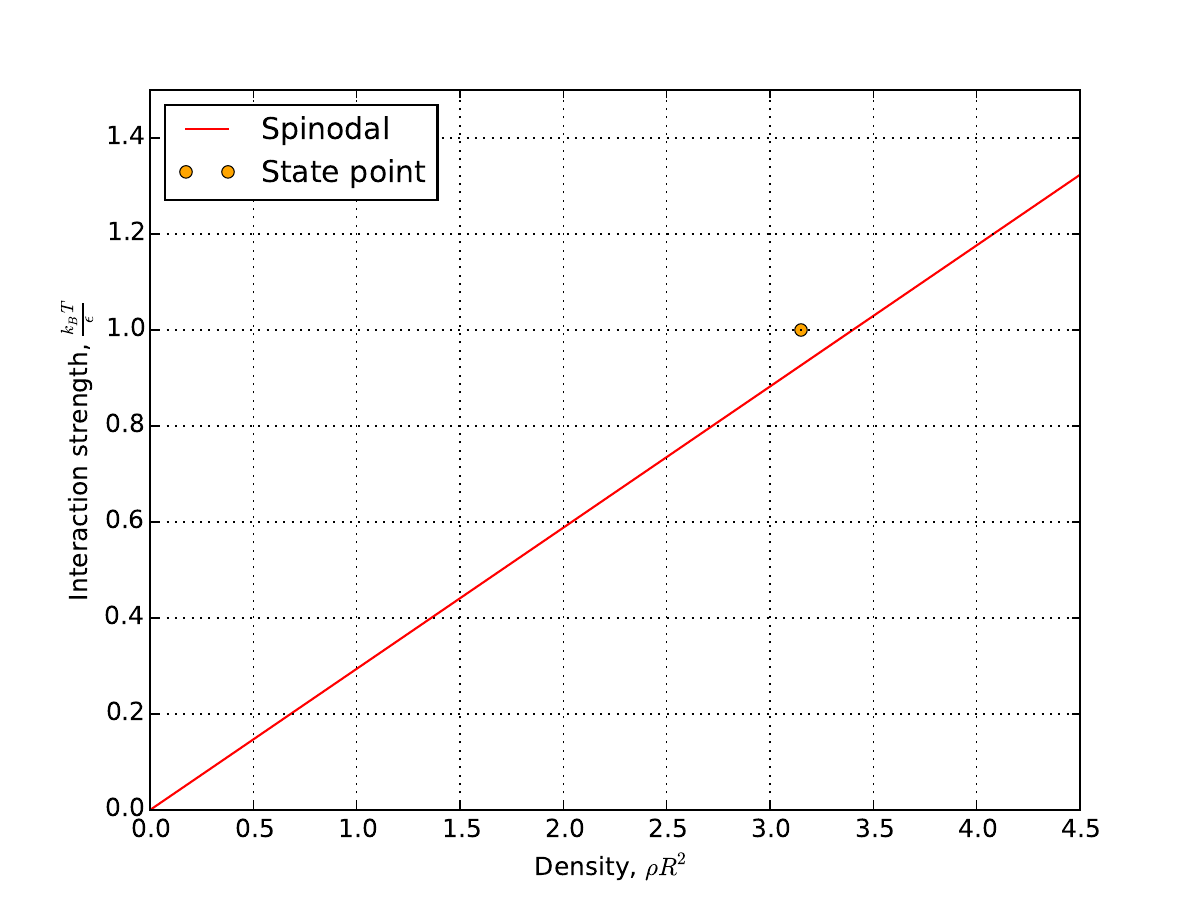}
\caption{The bulk spinodal line from mean-field theory for the GEM-8 model. The 
circle indicates the statepoint at which we perform our numerical calculations. 
On the left side of the line the system is a liquid and on the right it is in 
the crystal phase.}
\label{phase_diagram}
\end{figure}
The suspension is confined to a channel by the following external potential field
\begin{equation}\label{walls}
V_{\rm ext}(y)=\begin{cases} \;\;0\>\>\>\>  0<y<L_{y}  \\
\;\;\infty\>\>\> $otherwise$		
\end{cases},
\end{equation}
where $L_y$ is the distance between the walls. 
In a dynamical calculation the infinitely repulsive walls described by \eqref{walls} are 
equivalent to imposing a no-flux boundary condition. 
We choose the shear deformation such that the velocity field takes a Poiseuille form, with stick 
boundary conditions.
The velocity field is given by 
\begin{equation}
v(y)=4 v_m\left(\frac{y}{L_y}-\frac{y^2}{L_y^2}\right),
\end{equation}
where $v_m$ is the maximal velocity (which occurs in the center of the channel). 


\section{Theory}\label{theory}
The dynamical density functional theory (DDFT) provides a convenient approximate 
method to study 
the dynamical response of the particle number density, $\rho(\textbf{r},t)$, to  
external force fields.  
The original formulation of DDFT, which can be derived from either the Langevin 
\cite{marconi_tarazona_1, marconi_tarazona_2} or Smoluchowski \citep{archer_evans} 
descriptions, considered time--dependent external potential fields. 
Generalization to treat external flow fields (e.g.~shear) was first made by 
Rauscher {\it et al.} by incorporating an additional term representing the affine 
solvent flow field~\cite{rauscher}. 
However, it subsequently became apparent that this simple approach ignores important 
non--affine particle motion, which renders 
the approach incapable of describing interaction--induced currents orthogonal to the affine 
flow. 
As a consequence, the DDFT of Ref.~\cite{rauscher} cannot capture the physics of laning, 
particle migration or other phenomena arising from a coupling between flow and interparticle 
interactions.   

These shortcomings were addressed by Kr\"uger, Brader and Scacchi in Refs.~\cite{KB1,KB2,SKB}, who 
reintroduced the missing non-affine motion into DDFT using a dynamical mean--field approximation. 
In it's most recent form the DDFT equation is given by
\begin{equation}\label{ddft}
\frac{\partial \rho(\textbf{r},t)}{\partial t}+\nabla\cdot\left(\rho(\textbf{r},t)\textbf{v}(\textbf{r},t)\right)=\nabla\cdot\left(\Gamma\rho(\textbf{r},t)\nabla\frac{\delta 
\mathcal{F}[\rho(\textbf{r},t)]}{\delta\rho(\textbf{r},t)}\right)
\end{equation}
where $\textbf{v}(\textbf{r})$ is the solvent velocity field, $\Gamma$ is the mobility and 
$\mathcal{F}$ is the Helmholtz free energy containing details of the interparticle interactions and 
the external potential field. 

The Helmholtz free energy can be split into ideal, excess and external field contributions  
\begin{equation}\label{split}
\mathcal{F} = \mathcal{F^{\rm id}} + \mathcal{F^{\rm exc}_{\rm MF}} 
+ \int d\textbf{r}\, \rho(\textbf{r})V_{\rm ext}(\textbf{r}). 
\end{equation}
The ideal gas part is known exactly and is given by 
\begin{equation}
\mathcal{F}^{\rm id}[\,\rho\,] = k_{\rm B}T\! \int d\textbf{r}\, \rho(\textbf{r})\left(
\log\left(\rho(\textbf{r})\right) - 1	
\right),
\end{equation}
where we have set the thermal wavelength equal to unity. 
For the GEM-8 model presently under consideration the excess part can be well described by the 
mean--field functional
\begin{equation}
\mathcal{F}^{\rm exc}_{\rm MF}[\rho]=\frac{1}{2}\int d\textbf{r}\int d\textbf{r}'\rho(\textbf{r})\rho(\textbf{r}')\phi(\mid \textbf{r}-\textbf{r}'\mid).
\end{equation} 
The central approximation made here is that the two body density can be written in 
factorized form, 
i.e. $\rho^{(2)}(\textbf{r}, \textbf{r}')=\rho(\textbf{r})\rho(\textbf{r}')$. 
Despite the simplicity of this approximation, it has recently 
been pointed out that it performs better than one would expect \cite{chacko}.

The velocity field can be decomposed into an affine term and a fluctuation term,  according to
\begin{equation}
\textbf{v}(\textbf{r},t)=\textbf{v}^{\rm aff}(\textbf{r},t)+\textbf{v}^{\rm fl}(\textbf{r},t).
\end{equation}
Kr\"uger and Brader have shown \cite{KB1,KB2} that the fluctuation term in the velocity field can be 
approximated by the mean--field form
\begin{equation}\label{convolution}
\textbf{v}(\textbf{r},t)=\int\!d\textbf{r}'\rho(\textbf{r}',t)
\,\boldsymbol{\gamma}(\textbf{r}')\,\boldsymbol{\kappa}(\textbf{r}-\textbf{r}'),
\end{equation}
where $\kappa(\textbf{r})$ is a time-independent Kernel depending on the interparticle 
interaction potential and 
$\gamma(\textbf{r})$ is the position dependent shear flow. 
While the convolution form of the mean--field term is both physically plausible and computationally 
appealing, it nevertheless represents an empirical 
add-on to the microscopically derived original versions of DDFT 
\cite{marconi_tarazona_1, marconi_tarazona_2,archer_evans} - a true microscopic derivation 
would indeed be very useful. 
As a consequence the detailed form  of the convolution kernel $\boldsymbol{\kappa}$ remains unspecified and lacks 
a rigorous microscopic prescription. One can, however, give a physical interpretation to the kernel in terms of binary collision events, which facilitates the development of approximations.

\begin{figure}[t!]
\includegraphics[scale=0.54]{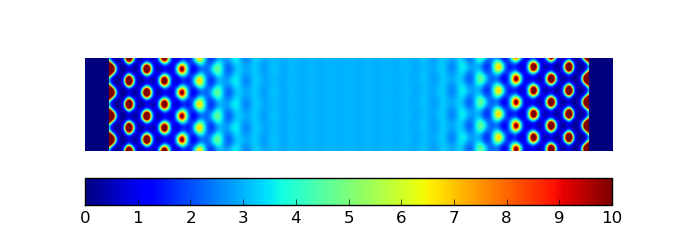}
\caption{Density of the equilibrium system. Close to the substrate we see a pre-crystal region, where the particles 
form a cluster crystal \cite{Archer_GEM}. 
In the center of the system we have a rather uniform density, with some residual oscillations due to packing effects.}\label{initial}
\end{figure}

The product of the kernel with the shear--rate 
in \eqref{convolution} has the units of velocity. This non-affine velocity arises from the 
force of interaction between a pair of particles when they are driven into each other by the 
flow field (`flow--interaction coupling') and thus encodes a non--affine motion. 
Physically, one can envisage `rolling over' type motion as two spherically symmetric particles 
attempt to follow the affine flow as closely as possible. 
Noting that for over damped systems the velocity and force are equivalent (up to a friction coefficient),  
a simple ansatz for the kernel is
\begin{equation}
\boldsymbol{\kappa}(\textbf{r})=-\alpha(\rho)\nabla\phi(\textbf{r}).
\end{equation}
In general one would expect the coefficient $\alpha$ to have a density dependence (not necessarily 
local). However, in the absence of detailed information about this function we choose the simplest 
possibility and set $\alpha=1$.  
We will see that this already leads to some interesting phenomenology and leaves open the possibility 
for fine tuning in order to fit data from numerical simulations or experiments. 


The equilibrium state corresponds to the long-time limit of equation \eqref{ddft} in the absence of flow. 
Equivalently, the equilibrium density can be obtained by minimizing the grand potential 
\begin{equation}
\Omega[\rho] = \mathcal{F}[\rho] - \mu\!\int d\textbf{r}\, \rho(\textbf{r}),
\end{equation} 
where $\mu$ is the chemical potential. 
The minimization generates the Euler-Lagrange equation
\begin{equation}\label{density}
\rho(\textbf{r})=\exp[\beta\mu-c^{(1)}(\textbf{r})-\beta V_{\rm ext}(\textbf{r})],
\end{equation}
where $\beta=1/k_B T$ and 
the one-body direct correlation function $c^{(1)}(\textbf{r})$ is given by a 
convolution with the interaction potential
\begin{equation}
c^{(1)}(\textbf{r})=-\beta\int\> d\textbf{r}'\rho(\textbf{r}')\phi(\mid\textbf{r}-\textbf{r}'\mid).
\end{equation}
The bulk phase diagram shown in Fig.~\ref{phase_diagram} has been calculated by solving 
Eq.~(\ref{density}) in the case of vanishing external potential. 
%
\begin{figure}[t!]
\includegraphics[scale=0.54]{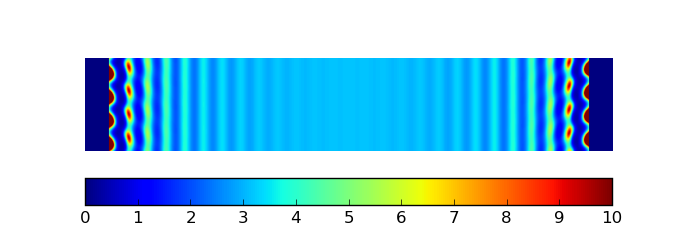}
\caption{Density at time $t=10$. 
The system has undergone a linearly increasing Poiseuille flow, reaching the maximal amplitude, 
$v_m=20$, at the moment shown here. The pre-crystal region at the substrates is less important, and has been replaced by smoother stripes.}\label{after_ramp}
\end{figure}
For densities above the spinodal the uniform liquid becomes linearly unstable with respect to the 
inhomogeneous crystal--phase density distribution.  
As previously mentioned, because of the soft--core nature of the particles the crystal phase 
consists of clusters of particles occupying the sites of a two-dimensional hexagonal lattice. 
The phase diagram shown in Fig.~\ref{phase_diagram} applies to an infinite system and we note that 
small discrepancies can occur in our numerical DDFT calculations arising from finite size effects. 
As the crystallising system is rather sensitive to the exact location of the statepoint we have 
been careful to keep to a minimum finite size effects arising from the confinement/periodic 
boundaries employed in our dynamical calculations.
For more details 
about boundary effects, we refer the reader to \cite{Archer_GEM}, where a detailed study of 
crystallisation under confinement at interfaces and in wedges is performed.

\begin{figure}[t!]
\includegraphics[scale=0.54]{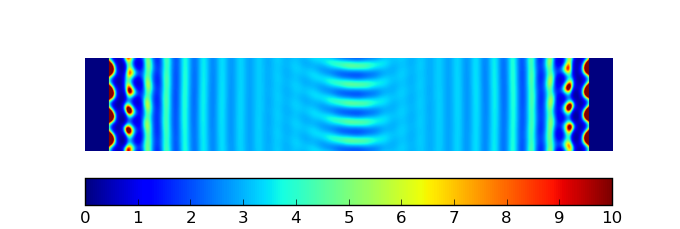}
\caption{For $t=37.5$. A symmetry breaking along the flow direction occurs at the center of the channel. 
The stripes which form which follow the parabolic flow profile.
}\label{first_break}
\end{figure}

\section{Results}
We prepare the system at a thermodynamically stable state point close to the crystal phase, as shown in 
Fig.~\ref{phase_diagram}. 
In Fig.~\ref{initial} we show the equilibrium (unsheared) density distribution of a system at this 
statepoint; average density $\bar{\rho}R^2=3.15$, and with size $(6R,31.1R)$. 
The average density is calculated by integrating over the density profile and dividing by the system 
volume.
Note that to improve visualization of the density profile we saturate the 
colorbar at a density value of $\rho R^2=10$.
This initial density distribution was calculated by solving equation \eqref{density} at chemical 
potential $\mu=9.7$. 
As we are working quite close to the freezing phase boundary there is clear evidence of 
pre-crystallisation at each wall. 
We choose to perform our calculations quite close to the phase boundary for two reasons: 
(i) it is interesting to see how switching on the Poiseuille flow will influence the 
pre-crystal regions at each wall and (ii) the flow--induced crystallisation phenomena we wish to 
investigate can be observed at relatively low shear rates;  
high shear rates generate stability issues which make difficult an accurate numerical solution 
of the DDFT equation.

The integration step used here is $dt=10^{-3}$. To avoid numerical issues, we switch on the flow using a ramp function for the first $10$ Brownian time units. After this time, the flow reaches is maximum value with $v_m=20$. At this moment, the system is only slightly distorted, with the majority of the distortion 
around the walls. 
The density peaks become smeared out and a stronger laning effect is visible over the entire system, 
see Fig.~\ref{after_ramp}. 

\begin{figure}[t!]
\includegraphics[scale=0.54]{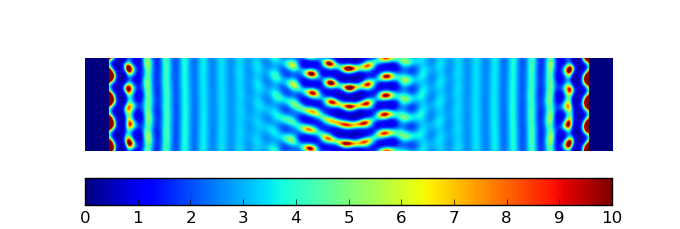}
\caption{For $t=40$. A second symmetry breaking occurs as the stripes destabilize to form 
density peaks and a crystal structure begins to develop.
}\label{second_break}
\end{figure}

From $t=10$ onwards the system undergoes a steady Poiseuille flow. 
For about further $25$ time units the density does not undergo any significant change. 
At $t=37.5$ (see Fig.~\ref{first_break}) the density distribution starts to change considerably as the migration mechanism 
pushes more particles to the center of the channel and the system undergoes a 
symmetry breaking along the flow direction in the central region of the channel.  
The `stripes' which emerge have a parabolic form similar to that of the external flow, but 
can only be 
observed within a relatively short time-window. 
These structures serve as a precursor to the formation of a more
crystalline region at the center of the channel.  
As time progresses each of the stripes develops some internal structure and distinct 
density peaks begin to emerge: Fig.~\ref{second_break} shows the situation at $t=40s$.

\begin{figure}[b!]
\includegraphics[scale=0.54]{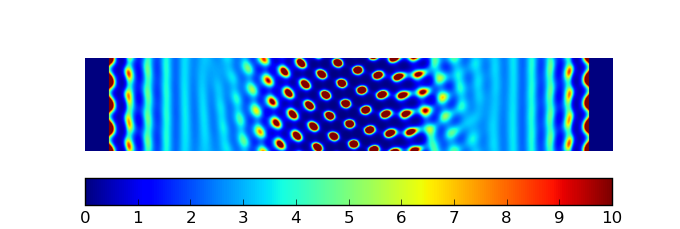}
\caption{For $t=50$. The system has entered a steady-state with a local crystal structure at the centre. 
As the system is under steady driving the crystal structure is constantly deformed and the peaks 
change their relative position as the density attempts to follow the affine Pousseuile flow.
}\label{crystal}
\end{figure}

Within approximately 10 time units the density peaks stabilize to form a steady--state and no further 
qualitative change in the density is observed. 
In Fig.~\ref{crystal} we show a representative steady--state density distribution at $t=50$. 
Because the system is under constant external drive the steady--state crystalline 
structure at the channel center is subject to continuous deformation; the density peaks 
try to follow the affine flow, but have to `squeeze past' their neighbours in order to do so. 
Despite this constant rearrangement a general hexagonal packing structure can be identified 
at any given time.  

\begin{figure}[t!]
\includegraphics[scale=0.46]{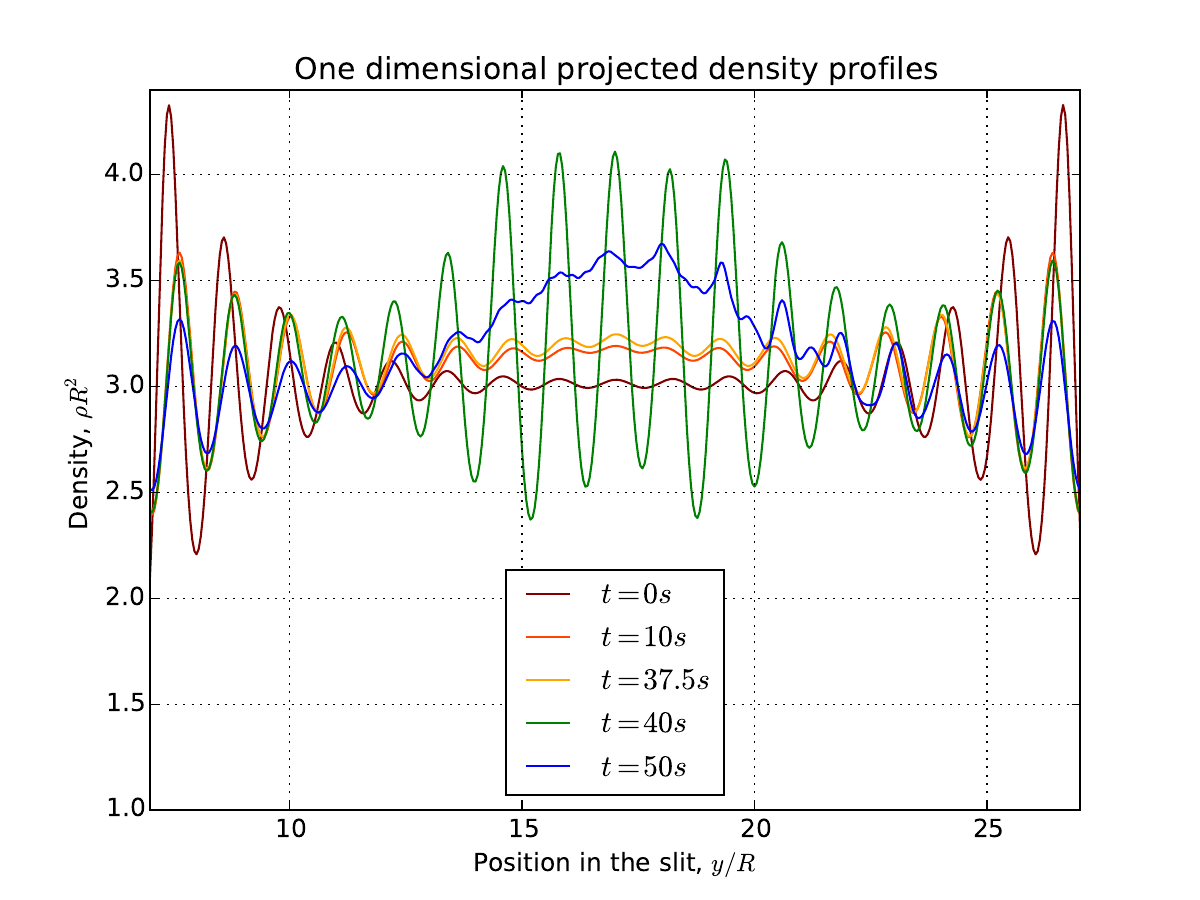}
\caption{Projected densities accross the channel corresponding to the different 
snapshots shown in the preceeding figures.}\label{fig_7}
\end{figure}

To provide an alternative visualisation of this phenomenon we show in Fig.~\ref{fig_7} one-dimensional 
density distributions obtained by integrating the two-dimensional density along the direction of 
flow. Data is shown for five different times to illustrate important stages 
of the time-evolution. 
%
Following the onset of flow the density gradually increases in the center of the system 
(from brown to yellow lines) as a consequence of shear-induced particle migration. 
Once the local density around the channel center approaches the bulk crystallization 
phase boundary density peaks start to form in the two-dimensional density distribution, which 
lead to the strong oscillations in the integrated data shown in Fig.~\ref{fig_7} for 
$t=40$ (c.f. Fig.~\ref{second_break}). %
For later times we enter a steady--state situation and the oscillations get smoothed out as a result 
of {averaging over the continuous distortion of the crystal structure when in steady--state.

Within the framework of DDFT the flow-induced structural changes discussed above can be related 
to changes in the free energy of the system. The fact that we can analyze dynamic phenomena using 
the equilibrium concept of a free energy is a 
consequence of the adiabatic approximation underlying the DDFT. 
%
%
%
%
\begin{figure}[t!]
\includegraphics[scale=0.46]{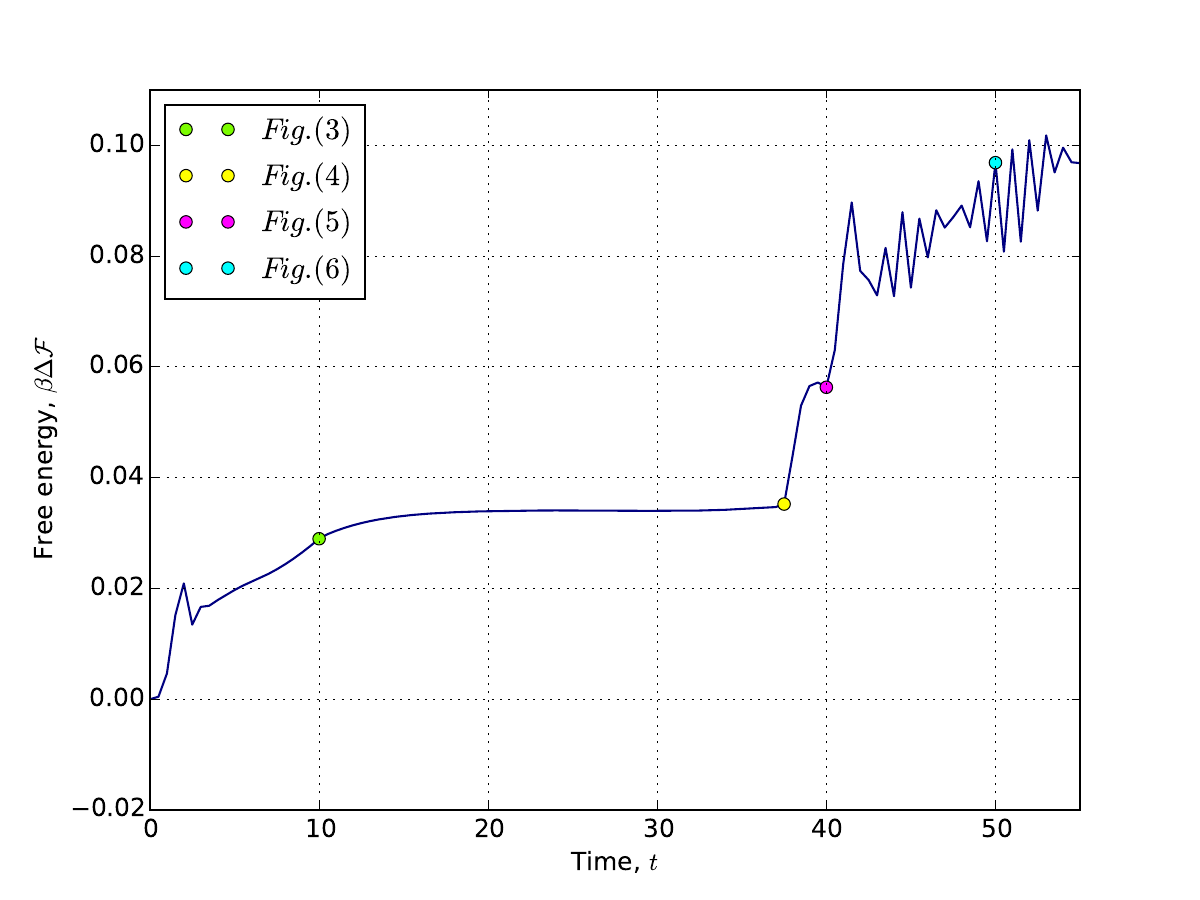}
\caption{Difference in the total free energy relative to that of the equilibrium state 
as a function of time. 
The points represent the free energy of the specific cases shown in figures 
3-6.}\label{fig_8}
\end{figure}
In Fig.~\ref{fig_8} we show the time-evolution of the Helmholtz free energy. 
For the driven system under consideration we recall that there is no `H-theorem' and that 
the free energy is not required to decrease on approach to the steady state; the applied shear 
flow is constantly adding energy to the system. 
The points in the figure indicate times for which the two-dimensional data is shown in the 
previous figures.
The free energy increases steadily up to around $t=20$ and then remains constant up to around 
$t=35$. 
Up until this point in time the density does not show significant change, other than the erosion of 
pre-crystal structures at each wall.
At around $t=37.5$ the first indication of a symmetry breaking occurs as the `stripes' start to appear 
in the center of the channel and this is reflected by a sudden increase in the free energy. 
At $t=40$ the stripes become unstable and density peaks develop, leading to another jump in 
free energy. 
For later times the local crystal is formed and the value of the free energy oscillates. 
These oscillations are due to the continual deformation of the crystal as the nonuniform flow 
pushes the central peaks at a larger velocity than those located to the left or right of the center. 
We note that the details of these oscillations are somewhat influenced by the finite system 
size employed in our calculations.

\section{Discussion}\label{discussion}

The DDFT is a well established method to calculate approximately the relaxation to equilibrium 
of the one-body density in overdamped systems. Its direct application to driven systems is problematic 
due to the neglegt of nonaffine particle motion; a phenomenological correction to the theory, 
as embodied by the flow kernel, must be introduced to account for these in a mean-field fashion. 
We have presented a new approximation for the nonaffine velocity field, which is appropriate for 
treating systems with soft, penetrable interparticle interactions.  

By applying our theory to the GEM-8 model under Poisseuile flow at statepoints close to the 
freezing transition we have investigated the interaction between crystallisation and shear-induced 
particle migration. The latter mechanism is not accounted for in standard DDFT and only enters 
as a result of our mean-field treatment of the nonaffine velocity. 
Following the onset of flow we observe that, after a period of transient dynamics, a local 
crystalline steady-state can form at the center of the channel. We anticipate that this 
phenomenon, which we have not seen reported elsewhere, will be generic for a wide class of soft 
particle models. Whether hard-spheres or similiar systems with strongly repulsive interactions 
would also exhibit this effect remains an open question. 

The present study presents much opportunity for development and future investigation. Of considerable 
interest would be stochastic simulations to establish the limitations of our phenomenological DDFT 
results. It would also be very interesting to investigate the impact of microstructural ordering 
(laning and local crystallization) on the rheology of the system. For example, one could enquire 
whether the transition to a nonequilibrium crystalline state increases the throughput in channel 
flow at a prescribed pressure. These and other questions will be the subject of further research.

On a more fundamental level, we are currently employing methods of bifurcation/stability analysis 
to investigate in more analytic detail the onset of symmetry breaking in `non-affine DDFT', 
i.e.~DDFT with a convolutional flow kernel term, such as that considered in the present work. 
We have so far focussed on the laning instability under simple shear flow 
\cite{instability}, but we anticipate that similar techniques could be fruitfully employed to 
study in detail, and with reduced numerical effort, more general flow induced crystallization 
and ordering phenomena. 
Work in this direction is currently in progress.

\section{Acknowledgement}
We thank Andrew J. Archer for interesting discussions and the Swiss National Science Foundation for financial support under the grant number 200021-153657/2.

\end{document}